\documentstyle[12pt,aaspp4,flushrt]{article}
\makeatletter

\@ifundefined{chapter}{\def\thebibliography#1{\section*{References\@mkboth
  {REFERENCES}{REFERENCES}}\list
  {\relax}{\setlength{\labelsep}{0em}
        \setlength{\itemindent}{-\bibhang}
        \setlength{\itemsep}{0pt}
        \setlength{\parsep}{0pt}
        \setlength{\leftmargin}{\bibhang}}
    \def\newblock{\hskip .11em plus .33em minus .07em}
    \sloppy\clubpenalty4000\widowpenalty4000
    \sfcode`\.=1000\relax}}%
{\def\thebibliography#1{\chapter*{Bibliography\@mkboth
  {BIBLIOGRAPHY}{BIBLIOGRAPHY}}\list
  {\relax}{\setlength{\labelsep}{0em}
        \setlength{\itemindent}{-\bibhang}
        \setlength{\itemsep}{0pt}
        \setlength{\parsep}{0pt}
        \setlength{\leftmargin}{\bibhang}}
    \def\newblock{\hskip .11em plus .33em minus .07em}
    \sloppy\clubpenalty4000\widowpenalty4000
    \sfcode`\.=1000\relax}}

\newlength{\bibhang}
\let\@internalcite\cite
\def\cite{\let\@citeleft(\let\@citeright)%
    \@ifstar{\citeyear}{\citefull}}
\def\acite{\let\@citeleft\relax\let\@citeright\relax%
    \@ifstar{\citeyear}{\acitefull}}
\def\citenp{\let\@citeleft\relax\let\@citeright\relax
    \@ifstar{\citeyear}{\citefull}}
\def\citefull{\def\astroncite##1##2{##1~##2}\@internalcite}
\def\citeyear{\def\astroncite##1##2{##2}\@internalcite}
\def\acitefull{\def\astroncite##1##2{##1~(##2)}\@internalcite}

\def\@citex[#1]#2{\if@filesw\immediate\write\@auxout{\string\citation{#2}}\fi
  \def\@citea{}\@cite{\@for\@citeb:=#2\do
    {\@citea\def\@citea{; }\@ifundefined
       {b@\@citeb}{{\bf ?}\@warning
       {Citation `\@citeb' on page \thepage \space undefined}}%
{\csname b@\@citeb\endcsname}}}{#1}}

\def\@cite#1#2{\@citeleft#1\if@tempswa , #2\fi\@citeright}
\def\@biblabel#1{}

\makeatother

\newcommand{\PSbox}[3]{\mbox{\rule{0in}{#3}\includegraphics{#1}\hspace{#2}}}
\newcommand{\FigNum}[1]{\unitlength 1pt \begin{picture}(55,10)(-400,35) 
                        \put(0,0){Figure #1}
                        \end{picture}}
\newcommand{\perval}[2]{{#1\mbox{$^{#2}$}}}
 
\newcommand{\persec}{\perval{s}{-1}\/}
\newcommand{\percm}{\mbox{$\cm^{-2}$}}

\newcommand{\ppm}{\mbox{$\pm$}}
\newcommand{\cgsflux}{\erg~\percm~\persec}
\newcommand{\cgslum}{\erg~\persec}

\def\etal{{et~al.}}

\newcommand{\nh}{\mbox{$N_{\rm H}$}}

\def\chisqr{\mbox{$\chi^2$}}

\newcommand{\ee}[1]{\mbox{$10^{#1}$}}
\newcommand{\tee}[1]{\mbox{$\times 10^{#1}$}}

\newcommand{\cm}{\mbox{$\rm\,cm$}}

\newcommand{\erg}{\mbox{$\rm\,erg$}\/}

\newcommand{\chandra}{{\em Chandra\/}}

\newcommand{\rxte}{{\em RXTE\/}}

\newcommand\psr{PSR~B1821$-$24}  
\newcommand\musec{$\mu$s}  

\newcommand\ms{{ms}}
\newcommand\bbb{B02}  
\newcommand\nancay{Nan\c cay}  
\newcommand{\psrp}{3.05~ms}
\newcommand{\nus}{\mbox{$\nu_{\rm s}$}}

\newcommand{\dmunit}{pc~\perval{cm}{-3}}
\newcommand{\cperksec}{c~\perval{ksec}{-1}}

\received{}
\revised{}
\slugcomment{}

\begin{document}

\title{Micro-second Timing of \psr\ with \chandra/HRC-S}
\author{
Robert E.\ Rutledge\altaffilmark{1}, 
Derek W.\ Fox\altaffilmark{1}, S.\ R.\ Kulkarni\altaffilmark{1}, Bryan A.\ Jacoby\altaffilmark{1}, 
I.\ Cognard\altaffilmark{2}, D.\~C.\~Backer\altaffilmark{3}, and Stephen
S.\ Murray\altaffilmark{4}}

\altaffiltext{1}{
Division of Physics, Mathematics and Astronomy, California Institute
of Technology, 
MS 130-33, Pasadena, CA 91125;
rutledge@tapir.caltech.edu, srk@astro.caltech.edu,
derekfox@astro.caltech.edu, baj@astro.caltech.edu}
\altaffiltext{2}{
Laboratoire de Physique et Chimie de l'Environnement - CNRS, UMR 6115 -
F-45071 Orleans Cedex2  -  France ; icognard@cnrs-orleans.fr } 

\altaffiltext{3}{
Astronomy Department and Radio Astronomy Laboratory, University of
California, Berkeley;  Berkeley, CA 94720-3411;
dbacker@astro.berkeley.edu 
}
\altaffiltext{4}{Harvard-Smithsonian Center for Astrophysics, 60
Garden Street, Cambridge, MA 02138; ssm@head-cfa.harvard.edu}

\begin{abstract}
We perform absolute timing of \psr\ in M28, using a 50 ksec
observation with \chandra/HRC-S.  We have obtained the highest
signal-to-noise X-ray pulsed lightcurve of this source to date,
detecting two X-ray pulses, as well as significant non-pulsed emission
-- a persistent X-ray flux which comprises 15\ppm3\% of the total
X-ray flux of the pulsar.  The Gaussian width of the sharp X-ray peak
is 34\ppm3~\musec\ in time, implying a size of the X-ray beam as it
crosses the line of sight of 4.0\ppm0.4 deg.  We find evidence for a
significant trailing component in both X-ray peaks of the pulse
profile.  Including three \rxte/PCA observations in our analysis, and
tying the phases together using a radio ephemeris obtained at \nancay,
we find the absolute phases in the X-ray wander with respect to this
radio ephemeris by up to 60~\musec, likely due to the variable
dispersion measure, which changes the pulse arrival time in the radio
band but not the X-ray band.  The present analysis makes clear that
study of pulsar timing noise properties in millisecond pulsars such as
\psr\ -- hitherto only studied at radio wavelengths, where variable
dispersion measure requires a significant correction -- can be studied
at X-ray wavelengths, where the effect of variable dispersion measure
is negligible.  We also examine the known uncertainties in the
absolute \chandra/HRC-S timing accuracy, which amount to \ppm
12~\musec.  We limit the amount of linear drift in the relative timing
accuracy of HRC-S to $<$3\tee{-10} s~\perval{s}{-1}.
\end{abstract}


\section{Introduction}

\psr\ is an isolated, 3.05 \ms\ pulsar, the first radio pulsar to be
discovered in a globular cluster \cite{lyne87}.  The object has been
frequently timed
\cite{lyne87,foster88,foster91,cognard96,backer97}. Due to its low
ecliptic latitude, it has also been used to study the distribution of
the solar coronal plasma (see for example \citenp{cognard96}).  \psr\
is the second brightest magnetospheric X-ray pulsar in the sky
\cite{becker99}.  The high X-ray luminosity and millisecond period
makes this pulsar a preferred target for the study of pulsar emission,
as well as an excellent target to tie the clocks of satellite-borne
X-ray detectors to UTC (\citenp{rots98}; hereafter, R98).

The radio profile consists of three components, the height and width
of which exhibit significant variation across the decimeter band
\cite{backer97}.  In the 800 MHz band, the two most prominent peaks
(RP1 and RP2, separated by 107.6\ppm1.4 degrees; \citenp{foster91})
have narrow widths (about 80\,$\mu$s). There is also a third broad
component (RP3), which follows RP2 in phase \cite{backer97}.  The
X-ray profile consists of two peaks (Saito et al. 1997).  The ASCA
clock was accurate to link the X-ray profile to the radio profile to
no better than 200\,\musec .  Using RXTE, R98 demonstrated that XP1,
the stronger component, arrived after RP1 by $62\pm 13\,\mu$s.

Recently, Becker \etal\ (\citenp*{becker02}; \bbb\ hereafter)
described \chandra/ACIS-S spectral observations of the field of M28
sensitive to an X-ray flux limit of 6\tee{30} \cgslum, finding 46
point sources.  It was found that \psr\ had a phase average spectrum
corresponding to $\alpha$=1.2 power-law with an equivalent Hydrogen
column density \nh=1.6\tee{21} \perval{cm}{-2}, confirming earlier
phase-average spectral analysis \cite{saito97}.

Here, we present a 50-ksec observation with the High Resolution Camera
(HRC) aboard {\em Chandra} obtained with 15.625$\,\mu$s temporal
resolution (``S'' mode). The principal goal of the observation was to
tie the {\em Chandra} clock to UTC at the accuracy of 10$\,\mu$s, to
describe the sources of uncertainty in the UTC time tie, and to
confirm this tie with observations in the radio band. In addition, we
present analysis of three \rxte\ observations.  Finally, we took
advantage of the imaging capability of the HRC and searched for
pulsations in the 8 (of 12) brightest X-ray sources in the field. 

The paper is organized as follows.  In \S~\ref{sec:radio} we describe
the radio observations performed at \nancay, from which the pulsar
ephemeris (uncorrected for a time-variable dispersion measure -- DM)
was derived.  As the DM toward \psr\ varies significantly, we discuss
the present knowledge of the DM as this affects our interpretation
of the X-ray pulsar observations.  In \S~\ref{sec:chandra} we describe
the \chandra\ HRC-S observations and X-ray timing analysis, for both
\psr\ and for other X-ray sources detected in the field.  In
\S~\ref{sec:rxte}, we describe RXTE/PCA timing analyses of \psr, which
confirm the relative timing of the pulsar.  In \S~\ref{sec:comp}, we
compare the \chandra\ and \rxte\ pulsed lightcurves, folded on the
\nancay\ radio ephemeris.  In \S~\ref{sec:conclusions}, we briefly
discuss the implications of these observational results, and conclude.

\section{Radio Observations at \nancay}
\label{sec:radio}

The radio timing observations were conducted at the large decimetric
radio telescope located near \nancay\ (France). The collecting area is
$\sim$7000 $m^{2}$, equivalent to a 93 meter dish.  The system
temperature was typically $\sim$45~Jy before the upgrade in 1999-2000,
after which it was improved to 25~Jy.  The maximum integration time of
this transit telescope is 70 minutes. Timing observations of \psr\ are
conducted 8 to 12 times a month at different radio frequencies.

At \nancay, the pulsar signal is de-dispersed by using a swept
frequency oscillator (around 80MHz) in the receiver IF chain, at a
frequency of $\sim$1410 MHz.  The pulse spectra are produced by a
digital autocorrelator with a frequency resolution between 6.25 and
25kHz, depending on the observing frequency. The station UT time scale
is provided by a local Rubidium Frequency standard.  The offset
relative to the international UTC time scale was measured daily at
14hUT via the Observatory of Paris by a special-purpose receiver using
TV signals until December 1995, with accuracy of \ppm40~ns.  After
December 1995, a GPS common view system with the Observatory of Paris
was adopted and an even better link was achieved.  With the \nancay\
Timing System, the frequency of arrival is measured by
cross-correlation of the integrated pulse spectrum with a pulse
template.  The template profiles are constructed by integration of
individual pulse profiles detected during our observations.  The
frequency of arrival measured for a start time of the \nancay\
dedisperser is dual of the times of arrival measured by most other
observatories. We will adopt the traditional term of Time Of Arrival
(TOA) hereafter since the data analysis is ultimately based on pairs
of times and frequencies of arrival.

The data have been analysed by our software AnTiOPE using the Jet
Propulsion Laboratory Ephemeris DE405 for the Earth orbital motion and
for the spatial reference frame and the conventional UTC time scale
for time reference. The small corrections UT1-UTC for the Earth
orientation have been applied.  However, no corrections have been made
for the time-variable DM in the direction of \psr, which can affect
the pulse TOA by as much as $\sim$50~\musec\  at 1410 MHz (see the
following section). More details on the analysis can be found in
\acite{cognard95}.

For JPL ephemeris DE405, we performed an analysis using pulsar TOAs
from 1996 through 2002.  Proper motion was set to zero and position
held constant.  We used as time of origin MJD 51468.0 (Oct 17, 1999).
Table~\ref{tab:ephem} gives the rotational frequency and its first two
derivatives, a time of origin ($t_0$) and the pulsar phase ($\phi$) at
$t_0$, where $\phi=0$ corresponds to the arrival of RP2, and has an
average value of zero over the duration of the ephemeris.  We refer to
this ephemeris as the \nancay\ ephemeris.

\subsection{Time-Dependent Dispersion Measure}

The DM along the line of sight to \psr\ produces a finite time delay
between photons with frequency $\nu_1$ and $\nu_2$ \cite{cognard96}:

$$
\delta {\rm DM} = \frac{\delta t_{\nu_1} - \delta t_{\nu_2}}{k}
\frac{\nu_1^2\nu_2^2}{\nu_2^2 - \nu_1^2}
$$

\noindent which gives a linear relation between a change in the pulse
TOA at 1410 MHz ($\delta T_{1410\rm \, MHz}$), compared with that at
X-ray frequency (effectively, infinite frequency), and a change in the
DM ($\delta {\rm DM}$) of

$$
\delta T_{1410\rm \,  MHz}=  2.09\times10^{-3} \delta {\rm DM } \,\, 
{\rm sec}
$$

\noindent Observations at radio wavelengths have found that the DM
toward \psr\ varies significantly on a year's timescale
\cite{backer93,cognard97}, with an time-average increase between 1989
Oct and 1996 Jan of 0.005 \dmunit~\perval{yr}{-1}, or
10~\musec~\perval{yr}{-1} at 1410 MHz; the increase in DM was not
monotonic, but appears to be the result of a power spectrum of
interstellar plasma turbulence \cite{cognard96}.  In 1996 Jan, the DM
was measured to be 119.847(2)~\dmunit.

Radio observations from \nancay\ at 1280, 1410 and 1680 MHz, taken
between Jan and Sept 2002, measured the DM.  A mean value of
119.873(5) pc~\perval{cm}{-3} was observed, across a range spanning
119.865--119.883 \ppm0.010~pc~\perval{cm}{-3}.  This is comparable to
the value expected by extrapolating the 1989-1996 time average
increase.  

Moreover, the increase in DM between 1996 Jan and the mean value for
Jan-Sep 2002 result in a $\delta T_{1410\rm \, MHz}$=54\ppm11~\musec.
\label{sec:dm}

\section{\chandra\ Observations and Analysis}
\label{sec:chandra}

Observations were performed with the \chandra/HRC-S detector
\cite{hrc1,hrc2}. The HRC-S has timing resolution of 15.625~\musec, a
nominal energy response of 0.1-2 keV, and essentially no energy
resolution.  The observation began 2002 Nov 8 5:41 UT, integrating for
49148~sec and a total livetime of 49068~sec (see Table~\ref{tab:obs}).
Data were analysed using CIAO v2.3 and CALDB v2.18. The absolute
astrometry of observation (from telescope pointing) was checked using
the present knowledge of its accuracy
\footnote{ CIAO thread
http://asc.harvard.edu/ciao/threads/arcsec\_correction/index.html\#calc\_corr};
no known astrometric problems were found, and the absolute pointing
accuracy is 0.6\arcsec, 1$\sigma$.   

We identify \psr\ near its \chandra\ pointing position; a
power-density spectrum of the X-ray source counts \cite{press} finds
the pulsation with the appropriate period ($\approx$327 Hz).  We used
\psr\ for absolute astrometry, adopting for its position the \nancay\
radio ephemeris position (see~\S~\ref{sec:radio}). 

We created an image using {\tt wavdetect}, first binning the raw HRC
image (0.13175\arcsec\ per pixel) by a factor of 2 and searching for
sources on scales of a factor of 1, 2 and 4 larger.  The resulting
reconstructed wavelet image is shown in Fig.~\ref{fig:wave}. We
detected 13 sources within 2.6\arcmin\ of the cluster center.  For
each source, the absolute positions, fixed relatively to the radio
position of \psr\, and X-ray countrates are given in
Table~\ref{tab:catalog}.

To obtain comparative background, we extracted counts from two
circular regions 21\arcsec\ in radius, set off from the cluster center
by 32\arcsec\ and 37\arcsec, which did not include any identified
point sources.  In these, we found a total of 8583 background counts,
which corresponds to a total of 9.7\ppm0.1 background counts in a
1\arcsec\ radius region.  

\bbb\ found an excess diffuse X-ray component associated with M28
which extends to the core radius, which they attribute to unresolved
point sources.  To examine the contribution to background from this
diffuse component, as well as from the extended point-spread function
of point sources inside the core, we examined the countrate in two
4.47\arcsec\ circles within the core radius, the boundaries of which
are $>$2\arcsec\ from any resolved point source.  These regions had
countrates of 11.5\ppm0.5 counts in a 1 arcsec radius region
integrated over the livetime of the observation, an excess of
1.8\ppm0.5 counts over the more distant background regions, confirming
the excess diffuse emission found by \bbb.

We assume a distance of 5.5 kpc to M28 \cite{harris96}.  Throughout,
we apply an energy correction factor of 1.42\tee{-11} \cgsflux\ or
5.18\tee{34} \cgslum\ (0.1-2.0 keV) per 1 HRC-S count/sec, which is
the X-ray flux corrected for absorption, assuming \nh=1.6\tee{21}
\perval{cm}{-2} and a power-law photon spectral slope of $\alpha=1.2$,
as observed for \psr\ (\bbb).  Alternatively, 1 HRC-S count/sec is
4.23\tee{-11} \cgsflux\ or 1.54\tee{35} \cgslum\ (0.5-8.0 keV),
unabsorbed, approximately a factor of 3.0 higher than in the 0.1-2.0
keV band.

We process the raw event times in the following manner
(see~\S~\ref{sec:times} for the detailed discussion of time
corrections and their uncertainties).  The event times delivered (that
is, which are given in the data files known as the event-1 and event-2
files) require two steps to place them in the JPL~DE405 reference
frame barycenter (TDB).  First, they are corrected for \chandra\
command-control-data-management propagation time (see
\S~\ref{sec:times}) by adding:

Corrected Event Time = Event Time + 281$\mu$sec

\noindent Second, we use {\tt axbary} to convert the corrected event
times to the JPL~DE405 reference frame barycenter (TDB), using the
level 1 \chandra\ ephemeris orbit file and the \nancay\ ephemeris
position. 

Counts from each source were extracted from within $\sim$2\arcsec\
circles around source positions.  The time-average luminosity of \psr,
found from the average HRC-S background subtracted countrate, is
(4.0\ppm0.2)\tee{32} \cgslum (0.1-2.0 keV), consistent with the
time-average value found by \bbb\ (1.28\ppm0.02\tee{33} \cgslum,
0.5-8.0 keV) after correcting for different passbands.

We performed a Fourier transform on the barycentered lightcurves, with
the time resolution 15.625 ~\musec , 630262500 time bins, and a
frequency resolution of 2.0308998\tee{-5} Hz using FFTW
\cite{fftw}. With the complex transform, we produced a
Leahy-normalized power density spectrum (PDS; \citenp{leahy83}) of
each source.

\subsection{\chandra/HRC-S Timing Analysis of \psr}

Examination of the pulsar harmonics in the PDS show the pulsation is
strongly detected up to the 16th harmonic.  To obtain the best-fit
frequency, we convolved the discrete Fourier transform window function
with the PDS within \ppm10 frequency bins of each harmonic for a trial
frequency, and mapped the sum of the convolution as a function of
trial frequency, with frequencies over-resolved by a factor of 32
(twice the number of harmonics), taking the peak of the curve to be
the best-fit frequency 327.40562430(5) Hz, consistent with that
determined from the radio ephemeris.  The change in frequency over the
course of our observation is small compared to our uncertainty.

After determining this frequency, we extracted data about the position
of \psr\ within a 1\arcsec\ circle; this decreases our background
counts by a factor of 4 over a 2\arcsec\ circle (from $\sim46$\ppm2 to
11.5\ppm0.5). The enclosed number of counts decreased from 446 to 396
counts, indicating a loss of $\sim$15.4\ppm0.5 source counts (4\%).
Using the \nancay\ ephemeris (Table~\ref{tab:ephem}) to determine the
relative pulsar phase at the time start time of the observation

$$
{\tt TSTART}=50814.0\, {  \rm  d (MJD)} + 153121000.174675 \, {\rm sec}
$$

\noindent which we found to be $\phi({\rm TSTART})=0.0617(11)$
periods.  We folded this data in time at the X-ray measured pulsar
frequency, and corrected the phase to the \nancay\ radio ephemeris.
This folded X-ray lightcurve is shown in Fig.~\ref{fig:phaselc} in
differential and integral form.


Two X-ray peaks are evident in the pulsed lightcurve, a ``sharp'' peak
near $\phi$=0.64 (XP1) and a ``dull'' peak near $\phi$=0.21 (XP2).
XP1 is associated in phase with the radio component RP1; R98 found XP1
followed RP1 at 800 MHz by 20\ppm4 milli-periods.  We refer to this
X-ray peak XP1.  XP2, however, has no obvious corresponding component
in the radio band.

XP1 has a maximum countrate, averaged over 30.54~\musec, of (67
counts)/(0.01$\times$49.068 ksec)= 135\ppm16 HRC-S counts~\perval{ksec}{-1},
corresponding to an intrinsic luminosity of (7.0\ppm0.7)\tee{33}
\cgslum (0.1-2.0 keV).  XP2 has an intrinsic peak luminosity of
(2.3\ppm0.4)\tee{33} \cgslum.

We examined counts outside of the main pulses to determine if they are
consistent with the number expected from background.  We summed all
the counts, excluding those within a phase of \ppm0.15 of XP1 and XP2,
finding a total of 29 counts, where we expect 4.6\ppm0.2 counts due to
background. This is a detection of excess counts with a probability of
$<$3\tee{-14} as being due to a fluctuation in the background
countrate.  Treating the excess counts as a constant intensity, the
time-average countrate is (29-4.6)/($0.40\times49.068$ ksec)=
1.2\ppm0.2 HRC-S counts~\perval{ksec}{-1}, which is 15\ppm3\% of the
time-average countrate of \psr.  The corresponding luminosity of the
unpulsed component is (6.2\ppm1)\tee{31} \cgsflux (0.1-2 keV).

\subsection{Timing Analysis for Other Sources in the Field}
 
Standard Fourier analysis searches \cite{press} are most appropriate
for the sinusoidal or near-sinusoidal pulsations seen from isolated
neutron star sources \cite{pzt99} and some X-ray binary pulsars
\cite{rudi98}.  In searching for X-ray pulsations from fast radio
pulsars, on the other hand, we anticipate a sharply-peaked pulse
profile and a correspondingly rich harmonic spectrum (c.f.\ Sec.\
3.2.2 of \citenp{rem02}). 

These expectations are borne out in the HRC-S dataset for \psr\
itself, which shows significant power out to the 23rd harmonic of its
fundamental \psrp\ period.  The maximum single-harmonic power is found
in the second harmonic, which has $\nu=2\nus$ and $P_2=404$ in the
\acite{leahy83} normalization, and the individual harmonic powers do
not drop below the 99.9\% confidence level until the 24th harmonic.
In a blind search for pulsations in this dataset, the
maximum-significance detection would come in a sum of the first 23
harmonics, which provide a total power of $P_{\rm 1-23}=1894$ and a
single-trial false-alarm probability of $1\times 10^{-367}$.

Searches for a pulsar weaker than \psr\ should be more conservative in
approach; reducing the observed harmonic powers of the pulsar by a
factor of ten results in a most-significant detection for the sum of
the first thirteen harmonics only, with the addition of further
harmonics weakening the detection.  We performed  a pulse search of the
strongest eight sources in the field, apart from \psr, summing the
first ten harmonics and covering the 0.001~Hz to 2~kHz frequency range
for the fundamental.  Even at the upper boundary of this frequency
range, our tenth harmonic frequency is less than the 32~kHz Nyquist
frequency of our observations.

We used the $Z^2$ search technique for photon event lists
\cite{buccheri+83}.  At full frequency resolution, this results in
roughly $10^8$ independent trials and should allow us to be sensitive
to summed powers of 95 or more, roughly 6.5\% the strength of the
signal for \psr.  We did not find any convincing candidate pulsations
in this sample.  In three cases, our maximum power exceeds the nominal
threshold of 95, but only slightly less-significant triggers at
unrelated frequencies are also apparent, which make the highest-power
triggers not credible.  We are not certain of the cause of this
anomalous power, but note that the HRC detector is subject to deadtime
effects, which will cause the photons from the pulsar \psr\ to impose
non-Poisson patterns on the data of other sources in the field.

\subsection{Comparison of Source Detections with the Results of \bbb}

In Fig.~\ref{fig:cr}a, we compare the HRC-S countrates with the ACIS-I
countrates for all sources detected (in both detectors) within
160\arcsec\ of the center of M28.  For the ACIS-I countrate, we summed
the countrates in the 0.2-1.0 and 1.0-2.0 keV bands listed by \bbb.
The different quantum efficiencies and background countrates make for
different detection limits for the two detectors.  For the sources
which were detected in both observations, the relation between
countrates in the two different detectors was approximately I(ACIS-I)
= (I(HRC-S)/0.4 c~\perval{ksec}{-1}) c~\perval{ksec}{-1}, with a
systematic uncertainty of a factor of 2.

Two sources detected in the HRC observation were not detected by \bbb\
(sources \#3 and \#12, using our numbering, Table~\ref{tab:catalog}).
The brighter (\#3), had an HRC countrate of 0.7\ppm0.1
c~\perval{ksec}{-1}, which would produce an ACIS-I countrate of 1.1
c~\perval{ksec}{-1} (\ppm 0.3 dex), well above the limit of $<$0.12
c~\perval{ksec}{-1}, if its spectrum were the same as the other
detected sources.  To account for the non-detection with ACIS-I, the
photon spectral slope of source \#3 would have to be $\alpha>9$, which
is far steeper than for any known X-ray source.  We suggest that
source \#3 is variable.  Source \#12 was detected with HRC-S at
0.3\ppm0.1 c~\perval{ksec}{-1}, which would produce an ACIS-I
countrate of 0.8 (\ppm 0.3 dex) c/~\perval{ksec}{-1}.

We did not detect 30 of the 41 sources detected by \bbb\ in our
160\arcsec\ image radius (undetected \bbb\ sources in order of
decreasing total countrate, using the numbering of B02: 28, 32, 30, 8,
14, 33, 21, 31, 16, 10, 20, 11, 6, 35, 39, 27, 24, 34, 7, 42, 9, 13,
40, 12, 18, 46, 43, 15, 44, 5).  All of these would have had HRC-S
countrates at or below the HRC-S detection limit, with the exception
of \bbb's source \#28, which was detected with ACIS-I at 3.2\ppm0.3
\cperksec, which would have produced 1.3 HRC-S \cperksec\ (\ppm 0.3 dex,
assuming a similar spectrum as other detected sources), instead of
$<$0.3 \cperksec.  Although this source was not found by {\tt wavdetect},
we extracted counts from within a 1\arcsec\ radius of the position
found by \bbb, and found a background subtracted number of counts of
16\ppm5, corresponding to 0.3\ppm0.1 \cperksec, indicating that the
source has faded by a factor $>$4.3\ppm2.2.

Figure \ref{fig:cr}b shows HRC-S countrate vs. the softness ratio
(0.2-1.0 keV/2.0-8.0 keV) for sources from \bbb\ (\bbb sources 5, 6, 7,
9, 12, 15, 18, 34, 35, 39, 40, 42, 43, 44, and 46 were excluded
because they have ACIS-I single-band countrates and/or softness ratios
consistent with zero).  The softness ratios of sources not detected by
the HRC-S observation have large relative uncertainties, but as a
group they are not spectrally harder than the sources we detected.
Thus, most of the non-detections in the HRC-S seem to be due to the
lower sensitivity of the HRC-S relative to the ACIS-I, most likely due
to the higher background countrate in the HRC-S detector.

\subsection{{\em Chandra} Timing Accuracy}
\label{sec:times}

The accuracy of the HRC event time tags (after correction for their
location in telemetry - see \citenp{murray02}) is limited by: (1)
digitization in the HRC time value; (2) the stability of the
spacecraft clock relative to ephemeris time; and (3) knowledge of the
absolute relation of the spacecraft time to UTC (that is, the
command-control-data-management (CCDM) propagation time).  The
magnitude of each of these are tabulated in
Table~\ref{tab:uncertainties}.

Event times for the HRC \cite{HRC96} are generated by measuring the
time of the event since the beginning of a reference time (called the
Science Header Frame, which occurs every eighth telemetry frame or
$8\times 0.25625=2.050$$\, $sec).  This relative event time is digitized
to 15.625~\musec\ due to the number of telemetry bits available (18
bits available to count the spacecraft interface unit clock that runs
at 1.024~MHz or 0.9765625~\musec~\perval{count}{-1} so the HRC time
tags are in units of 16 counts of this 1.024~MHz clock {[}=
15.626~\musec{]} to span the 2.050 sec interval). Thus the
digitization uncertainty in HRC times is, on average, 5~\musec.

The spacecraft clock which governs the telemetry is based on the use
of an ultra-stable oscillator (USO), running at a nominal frequency of
8.192~MHz, on board the observatory, with periodic time tagging of the
spacecraft clock at the receiving ground stations of the Deep Space
network (DSN).  Typically, the Chandra Observatory will have a DSN
contact for about 1 hour every 8 hours, during which real-time
telemetry is received at the ground stations (as well as Solid State
Recorder dumps). The real time data (usually 30 minutes or more) are
time tagged with the antenna arrival time providing a series of
correlations between the spacecraft clock counter (which has a nominal
period of 0.25625 sec/count) and JPL ephemeris time. The Chandra
Science Center Flight Operations Team (FOT) \cite{Watson02} fit these
correlations to a simple second order polynomial, and the residuals
are analyzed to provide a measure of the stability of the clock. The
typical rms residual measured from these data is 5~\musec\ or less.
The measurements show occasional excursions to 10$\, $$\mu $sec;
monitoring of these correlations over the operational lifetime of the
observatory has detected no larger anomalies in the spacecraft clock
stability. There is an observed secular trend in the clock period
(0.25625000905 sec/count to 0.25625001046 sec/count over a 1016 day
interval), which is due to cumulative radiation dose to the
electronics, and is fully compensated for through the correlation
function.  Thus, the uncertainty in the time-tag due to the stability
of the spacecraft clock relative to ephemeris time is \ppm5~\musec.

The knowledge of the absolute time for Chandra data comes from the
time tagging of data at the receiving (ground) antenna, and a
knowledge of all of the propagation time delays between the spacecraft
and the antenna, and any delays at the ground station between the UTC
clock and the time tagging. The FOT has determined the sum of all of
the systematic delays -- including the delay on the spacecraft -- of
281.25~\musec\ \cite{FOT02} which is the propagation time of the
spacecraft clock though the various elements of the spacecraft
command, control, and data management system (CCDM). These systematic
delay times are believed to be accurate to $\pm $10~\musec.  This
uncertainty includes the resolution of the UTC clock at the antenna,
errors in propagation delays due to orbit errors, errors in the CCDM
propagation time, and errors in ground propagation times.

In the case of the Chandra science instrument time tagging, the CCDM
propagation time is incorrectly applied to the event times during
pipeline data processing and must be removed from the distributed
event times (that is, $T_{\rm correct}=T_{\rm pipeline}+281$~\musec, for both
the Level 1 and Level 2 event files). However, this processing error
is totally systematic and adds no additional uncertainty to the
absolute time once it is corrected. Thus the overall uncertainty in
absolute event time is estimated by assuming the three errors are
uncorrelated and can be root square summed to yield $\sigma
=\sqrt{5^{2}+10^{2}+5^{2}}=12$~\musec.  Of these three sources of
uncertainty, only the 10~\musec\ CCDM propagation time can -- in
principle -- be better characterized and eliminated as a source of
uncertainty.  Thus, future CCDM characterization may ultimately
decrease the time-tag uncertainty to \ppm7~\musec.

\section{\rxte/PCA Observations and Analysis}
\label{sec:rxte} 

We analysed all three observations performed with the RXTE/PCA as
listed in Table~\ref{tab:obs}, the first of which (Sept 1996) has
previously been analysed (R98).  During the Sept 1996 and Feb 1997
observations, all five PCU units were operating, while during the Nov
1999 observation, only three of the five PCU units were operating.  We
used FTOOLS v5.1A (release 16 Jan 2002).   All data were obtained in
Good Xenon mode, which has time resolution of $2^{-20}$sec ($\sim1\,
\mu$sec).  We selected events detected in the first detector layer
only with an approximate energy range of 2--24 keV, selecting all
pulse height analyser (PHA) channels beginning at channel 0 (where the
instrumental cut-off is 2 keV) and up to 24 keV, depending on the
detector voltage settings.

We barycentered the events into the DE405 time system using {\tt
axbary} in the CIAO release used for the \chandra\ analysis; this
applies the 16~$\mu$sec intrinsic delay of the PCA instrument, the
\rxte\ fine clock corrections, and the addition of the {\tt TIMEZERO}
keyparameter.  The absolute systematic uncertainty of the RXTE/PCA
event times in TDB is estimated to be \ppm5~\musec\ after MJD 50567
(1997 Apr 29 UT), and \ppm8~\musec\ prior to MJD
50567\footnote{http://heasarc.gsfc.nasa.gov/docs/xte/time\_news.html}
(Table~\ref{tab:uncertainties}). This level of accuracy is established
by calibration several times daily by the RXTE Mission Operations
Center, ultimately limited by the accuracy of the ground clock used
for this calibration at White Sands
\footnote{http://heasarc.gsfc.nasa.gov/docs/xte/abc/time.html}.  A
detailed discussion of this timing calibration is given in R98. We
used the \nancay\ ephemeris to calculate the phase of each event, and
produced a folded pulsar lightcurve.  In each, the background
countrate is significant, such that the persistent countrate for the
pulsar is well below the detection limit.

\section{Comparison of Observations in the \nancay\  Ephemeris} 
\label{sec:comp}

With the \chandra\ and \rxte\ folded lightcurves tied to the \nancay\
ephemeris, we compare the relative phases.

In Fig.~\ref{fig:allphases}, we show the three RXTE/PCA and
\chandra/HRC-S lightcurves folded on the \nancay\ ephemeris. We fit
Gaussians to XP1 and XP2, finding the phase relative to the \nancay\
ephemeris for XP1 ($\phi_1$) and XP2 ($\phi_2$) and the width
($\sigma_1$ and $\sigma_2$).  For Obs. 1, due to the low
signal-to-noise, we held $\sigma_1$ and $\sigma_2$ fixed at the
average values of obs.\ 2 and 4, to obtain a best-fit in relative
phase.  We list these values in Table~\ref{tab:phases}, with the {\em
statistical} uncertainty in phase (not including the systematic
uncertainties of \ppm8~\musec\ (obs.\ 1), \ppm5~\musec\ (obs.\ 2 and
3) and \ppm14~\musec\ (obs.\ 4)).

We calculated the shift in the phases of each component for each
observation N, relative to the phases measured from obs.\ 2
(($\phi_{1, \rm obs 2}-\phi_{1, \rm obs N})/$(327.4056~Hz) for XP1 and
($\phi_{2, \rm obs2}-\phi_{2, \rm obs N})/$(327.4056~Hz) for XP2).  The
relative shifts in the phase of the two components in each observation
are consistent, though typically with large uncertainties.  However,
we find significant phase shifts in obs.\ 3 and 4 relative to the
radio ephemeris prediction and the observed phases of obs.\ 1 and 2. 

To produce a tighter constraint on the relative phases of each
observation, we calculated a cross-correlation function \cite{press}
using a template pulse form, and taking obs.\ 2 as $\delta\phi=0$.  We
used each of the pulsed lightcurves binned in phase with bin sizes of
0.01 phases (30.54 ~\musec). We then calculated

$$
CCF(j) = \sum_{i=1,n} I_i \times \left(\sum_{l=1,2}
\frac{A_l}{\sigma_l}\exp{ -\frac{((i-j)/n - \phi_l)^2}{2\sigma_l^2}} \right)
$$

\noindent where $i$ is the phase bin number, $I_i$ is the number of
counts in phase bin $i$, $l$ is the pulse component, $A_l$ is the
relative normalization of the pulse component, $\sigma_l$ is the
Gaussian width, and $\phi_l$ is the centroid of the pulse component.
The values of $\sigma_l$ were taken as an average of all four observed
values; $\phi_1$ is was taken as the best-fit value from obs.\ 2;
$\phi_1-\phi_2$=162.5\ppm0.7 $\deg$ was found by taking a weighted
average of all four values; and $A_l$ are the best-fit values from the
\chandra\ data (see Table~\ref{tab:pulselc}).  We fit a quadratic
function to the peak of the CCF (taking only the highest five values),
to determine the best-fit phase lag, which we give (as a time lag, the
X-ray $\Delta T_{obs=2,N}$ -- the relative time lag of obs.\ $N$ from
obs.\ 2) in Table~\ref{tab:phases}, along with the statistical and
systematic uncertainties.

\subsection{Width of XP1, and the HRC-S Timing Precision}
\label{sec:steve}

The HRC-S observation resolves XP1 with a Gaussian width
$\sigma_1=$0.0115(9) in phase, or 35~\musec\ (Table~\ref{tab:phases}),
subtracting the digitization time in quadrature, this results in a
width of 34\ppm3 ~\musec. 

We examined phases as a function of time since the observation start
of all counts which arrived at phase $\phi_1$ within a range of
\ppm2$\sigma_1$ (we found 164 such counts).  We found no obvious
systematic deviations in the phase as a function of time, indicating
that the width of this component is intrinsic to the pulsar, and not
due to a systematic wandering of the absolute timing calibration.
A fit of a constant phase using a $\chi^2$ minimization technique
\cite{bevington}, assuming an uncertainty in phase of 0.0115 (the
width of XP1) found a best-fit constant phase to be an acceptable
model ($\chisqr=127$, 163 degrees of freedom) for the phases of these
counts.

Fitting a linear model for the mean phase as a function of time (best
fit: $\chisqr=126$ for 162 dof) finds that the change in mean phase as
a function of time is $<$6.3\tee{-7} radians per second over the
duration of the observation.  This places a limit of $<$3\tee{-10}
s~\perval{s}{-1} (90\% confidence) on a linear drift in the relative
timing calibration of the HRC-S instrument.  

\subsection{Trailing X-ray Emission}

Examination of Fig.~\ref{fig:allphases} shows that the X-ray pulses
are asymmetric, with greater emission trailing the main pulse than
leading it. In XP1 and XP2, the number of counts in the phase range
[$\phi_{N}+2\sigma_N$, $\phi_{N}+0.15$] is 39 and 25, respectively;
prior to the pulse, the number of counts in the phase range
[$\phi_{N}-0.15$, $\phi_{N}-2\sigma_N$] is 13 and 7, respectively.  The
probability of observing 39 counts when 13 are expected is 5\tee{-9};
and of observing 25 counts when 7 are expected is 1\tee{-7}. 

\subsection{Radio and X-ray Pulsar Timing Noise} 
\label{sec:timingnoise}

In Fig.~\ref{fig:timingnoise}, we show the history of the pulsar
timing noise of \psr\ between 1996-2002, relative to the \nancay\
ephemeris.  The radio $O-C$ values have the sign convention that, if
the observed pulse arrives before the calculated value, then $O-C$ is
negative.  Deviations in the pulse arrival time of a magnitude as
great as \ppm(50-60)~\musec\ have been observed due to a combination
of pulsar timing noise and unmodelled propogation effects such as DM
variations.

For comparison between the radio $O-C$ and the X-ray $\Delta T$, we
estimated the radio $O-C$ value, by taking an average of the five
nearest (in time) radio observations, and using the scatter of the
value as the uncertainty (listed in Table~\ref{tab:phases}).  For
comparison with the X-ray, we then set all values of the radio $O-C$
taken at the time of the four X-ray observations to be relative to
that measured for obs.\ 2.  We show the derived O-C values in
Fig.~\ref{fig:timingnoise}.  While the X-ray and radio O-C values
match in observations \#1 and 2, they are discrepant in observations
\#3 and 4.

In Fig.~\ref{fig:money}, we plot the pulsar time delay in the radio
band relative to that observed in obs.\ 2 as a function of the time
delay in the X-ray band relative to that observed in obs.\ 2; thus, the
value corresponding to obs.\ 2 lies at the origin.  The two
observations (\#3 and 4) are consistent with an absolute offset of the
X-ray time delay from the radio time delay, found by a $\chi^2$
best-fit assuming an X-ray/radio delay slope of 1 and neglecting the
uncertainty in the radio phase:

$$
{\rm X-ray}\;  \Delta T= {\rm Radio (O-C)} - 51\pm7 \; \mu{\rm sec}
$$ 

A direct offset in the TOA phase of this magnitude can be explained by
an increase in DM between observation 2 and 3 of 0.023~\dmunit, which
is consistent with observations (see \S~\ref{sec:dm}).  

The uncertainty in the position and proper motion of the pulsar will
not affect the relative X-ray/radio phase offset, such as that we see
in Fig.~\ref{fig:money}.  An incorrect geometric assumption which
results in a deviation in the radio pulse phase relative to the
baseline ephemeris will also result in a deviation in the X-ray pulse
phase measured at the same time, since a deviation in geometry affects
only light travel time, affecting the X-ray and radio bands equally,
which is not what we observe.

\subsubsection{The Absolute Phase of X-ray Component 1}

In the \nancay\ ephemeris, the phase $\phi=0$ corresponds to an
average phase over the duration of the ephemeris of zero for RP2.
Assuming that $\phi_{\rm radio,2}-\phi_{\rm radio,1}$=0.299(4)
\cite{foster91}, and taking the $\phi_{\rm X-ray, 1 (R98)}-\phi_{\rm radio,
1}$=0.020(4) from R98, we can find the difference between the presently
measured phase of XP1 and that from the phasing of R98: 

\begin{equation}
\phi_{\rm X-ray, 1} - \phi_{\rm X-ray, 1 (R98)} = \phi_{\rm X-ray,
 1} - \left[ 1 + \frac{{\rm Radio}(O-C)}{3.054 {\rm ms}} - 0.299(4)\right] - 0.020(4)
\end{equation}

\noindent Taking values from Table~\ref{tab:phases}, we find that for
observations 1-4, this difference amounts to $-180\pm20$~\musec,
$-180\pm20$~\musec, $-$230\ppm20~\musec, and $-$250\ppm20~\musec,
respectively.  Thus, all four measured values of the X-ray phases for
XP1 in the present work are in disagreement with that found by R98,
including obs.\ 1 (the same data as analysed by R98). 

It is difficult to see how the present measurements can be resolved
with that of R98.  The relative phase of XP1 is similar in all three
RXTE observations (within 60~\musec) to that of the \chandra\ phase,
within the \nancay\ ephemeris -- a smaller variation than the
$\sim$200~\musec\ discrepancy.  As we find similar phases in both the
\chandra/HRC-S observation and the three RXTE/PCA observations, it
does not appear to be due to a satellite calibration issue.  The
discrepancy may be due in part to the differing radio ephemeris of
\nancay\ with that used by R98 (found using Green Bank and Jodrell
Bank Telescopes and a different observational techniques, in a
different radio passband).  DM corrections were applied to the R98
epehemeris, although not to the present ephemeris; however, the
variable DM can amount to at most $\sim$50~\musec\ delay, which is
smaller than the $-180$~\musec\ to $-250$~\musec\ difference we see
here.  

This issue can be examined more closely with an analysis which ties
the \nancay\ ephemeris to observations at Green Bank, which is
underway presently (Backer \& Cognard, in progress).

\section{Discussion and Conclusions}
\label{sec:conclusions}

We have performed fast timing of \psr\ with \chandra/HRC-S, resolving
the sharpest X-ray peak XP1 with a width 34\ppm3~\musec. This also
serves as a limit on a random timing calibration uncertainty variation
over this 50 ksec observation.  In addition, we place a limit of
$<$3\tee{-10} s~\perval{s}{-1} (90\% confidence) on a linear-drift in
the relative timing calibration of the HRC-S.  Moreover, the Gaussian
width of this component limits the width of the X-ray pulse beam as it
passes across the line of sight to be 4.0\ppm0.4 degrees.

We detect for the first time persistent X-ray emission in \psr, at a
level of 15\ppm3\% of the time-average X-ray flux, assuming identical
spectra for the pulsed and non-pulsed emission.  We also find trailing
X-ray emission, significantly in excess of the emission leading both
pulse components.  

The absolute tie of the HRC-S instrument to TDB/UTC time standard
compared with the known \ppm5 ~\musec\ capability of \rxte\ is
confirmed with the precision limited by the alignment of the XP1
relative to the \nancay\ ephemeris of the HRC-S and three \rxte/PCA
datasets.  We observe that the four pulse lightcurves are
significantly mis-aligned, offset by as much as 67~~\musec.  This
mis-alignment appears to be due to the time-variable DM in the
direction of \psr\ over the 1996-2002 period, during which the \rxte\
and \chandra\ observations were made.  The known time-dependent DM,
measured elsewhere, results in a change in the TOA of the radio pulse
-- but not in the X-ray pulse -- of the order of that observed. 

A correction to the radio pulse TOA can be applied to the \nancay\
data, using DM measurements obtained at \nancay\ in the radio band
taken from 1996-2002.  This analysis is under way (Backer and Cognard,
in progress).  Once this correction is applied, the radio pulsed
lightcurve data can be aligned absolutely in phase with the X-ray
pulsed lightcurve, which can confirm {\em in situ} the \chandra/HRC-S
absolute timing capability to the accuracy of the known uncertainties
in the RXTE timing (\ppm 5~\musec) and the \chandra/HRC-S timing (\ppm
12~\musec).  
 
Pulsar timing noise has yet to be measured in the X-ray band.  The
present observations demonstrate that the technical capability exists
to do so for \psr\ in the presently operating X-ray observatories
\rxte\ and \chandra.  Repeated X-ray timing observations of \psr\ would
produce a pulse-ephemeris independent of the effects of a time-varying
DM, without having to measure the DM for each observation as required
for this and other millisecond pulsars (such as PSR~1951+32,
\citenp{foster90}; PSR~1821+05, \citenp{frail91}; PSR~1557-50,
\citenp{deshpande92}; PSR 1855+09, PSR 1937+21,
\citenp{backer93,cognard95}).  Thus, the X-ray band provides the
capability of measuring pulsar timing noise without a time-variable DM
correction.  Such observations would provide confirmation of the
attribution of pulsar timing noise to spin-down processes, rather than
for example, to processes associated with magnetospheric emission.

\acknowledgements

This work was supported by the \chandra\ Guest Observer program. The
\nancay\ Radio Observatory is part of the Observatoire de Paris,
associated to the French Centre National de la Recherche Scientifique
(CNRS). The \nancay\  Observatory also gratefully acknowledges the
financial support of the Conseil Regional of the Region Centre in
France.

\clearpage

\begin{figure}[htb]
\caption{ \label{fig:wave} Wavelet reconstructed image of
\chandra/HRC-S observation centered on M28 (cross), showing the
locations of X-ray sources detected within the half-mass radius (outer
circle, 93.6\arcsec radius).  The small circle marks the 14.4\arcsec\
core radius. The cross marks the optical cluster center
(Harris 1996). The 11.27\arcmin\ tidal-radius is much larger than
the image. Sources 8, 9, and 10 are not included in this figure, as they are
well outside the half-mass radius (their positions are listed in Table 3).
 }
\end{figure}
\nocite{harris96} 

\begin{figure}[htb]
\caption{ \label{fig:phaselc} The Chandra HRC-S phase-binned
lightcurve of \psr, folded on the pulsar period ($\sim$3.05 msec), and
aligned to the absolute phase of \nancay\ ephemeris. There are
9.7\ppm0.1 background counts in this lightcurve (total). The
counts detected outside the main pulse are significantly above
background, indicating pulsar bridge emission at the level of
15\ppm3\% of the total X-ray flux (0.1-2.0 keV)}
\end{figure}

\begin{figure}[htb]
\caption{ \label{fig:cr} {\bf Panel (a):} The HRC-S countrate
vs. ACIS-I countrate (0.2-2 keV; from \bbb), for sources detected
within 160\arcsec\ of the center of M28.  Down to a countrate limit of
2 ACIS-I \cperksec, all sources are detected in the HRC-S with the
exception of \bbb\ source \#28 at 3.1 ACIS-I \cperksec, but $<$0.3 HRC-S
\cperksec.  Down to a countrate limit of 0.4 HRC-S \cperksec, all sources
are detected with ACIS-I, with the exception of our source \#3 at 0.7
HRC-S \cperksec, but $<$0.12 ACIS-I \cperksec.  (N.B. the ACIS-I countrate
upper-limits are estimated from the $<0.2$ \cperksec upper limit for the
full 0.2-8 keV energy range, decreased by $\sim$60\%, assuming a
$\alpha$=1.4 photon power-law spectrum). {\bf Panel (b):} The HRC-S
countrate vs. the ACIS-I softness ratio (0.2-1.0 keV/2.0-8.0 keV).
The non-detections in the HRC-S of sources detected with the ACIS-I do
not appear to be correlated with spectrum.  }
\end{figure}

\begin{figure}[htb]
\caption{ \label{fig:allphases} The pulsar lightcurves observed (from
top to bottom) with RXTE/PCA (1996 Sept, 1997 Feb, and 1999 Nov), and
\chandra/HRC-S (2002 Nov), folded on the \nancay\ ephemeris.  The
component number 1 and 4 (XP1 and XP2) are labeled in the second panel
from the top.  Note that XP1, near $\phi$=0.65, do not align in the
\nancay\ ephemeris, likely due to the absence of DM-variability
corrections in the \nancay\ ephemeris.   }
\end{figure}

\begin{figure}[htb]
\caption{ \label{fig:timingnoise} The Observed minus Calculated (O-C)
zero-phase arrival time (of RP2) of \psr\ using the \nancay\
ephemeris, as a function of time (small dots).  The top axis is marked
in UT years, the bottom axis in MJD.  Deviations as great as
67~\musec\ have been observed.  The $O-C$ value for the \nancay\ radio
observations at the time of the X-ray observations (open squares) and
for the four X-ray observations (crosses).  The radio data are given
in absolute phase (uncorrected for changes in DM), whereas the X-ray
points are taken relative to the phase of Observation \#2 (MJD 50489).
The X-ray points for observations \#3 and \#4 are discrepant with the
O-C values found from radio data.  The magnitude and sign of the
discrepancy is consistent with that expected from the observed change
in DM between 1996 and 2002. }
\end{figure}

\begin{figure}[htb]
\caption{ \label{fig:money} The X-ray pulse time delay, as a function
of radio pulse time delay, both taken relative to obs.\ 2 (which
therefore lies at the origin).  The uncertainty in the values for
obs.\ 2 are shown on that data point, although this uncertainty is
also reflected in the uncertainties of the other points.  The values
for obs.\ 3 and 4 lie clearly away from a relationship in which the
pulsar timing noise is of equal magnitude in both the radio and X-ray
bands.  This discrepancy can be explained as due the variable DM (see
\S~\ref{sec:dm}).  }
\end{figure}

\clearpage
\pagestyle{empty}
\begin{figure}[htb]
\PSbox{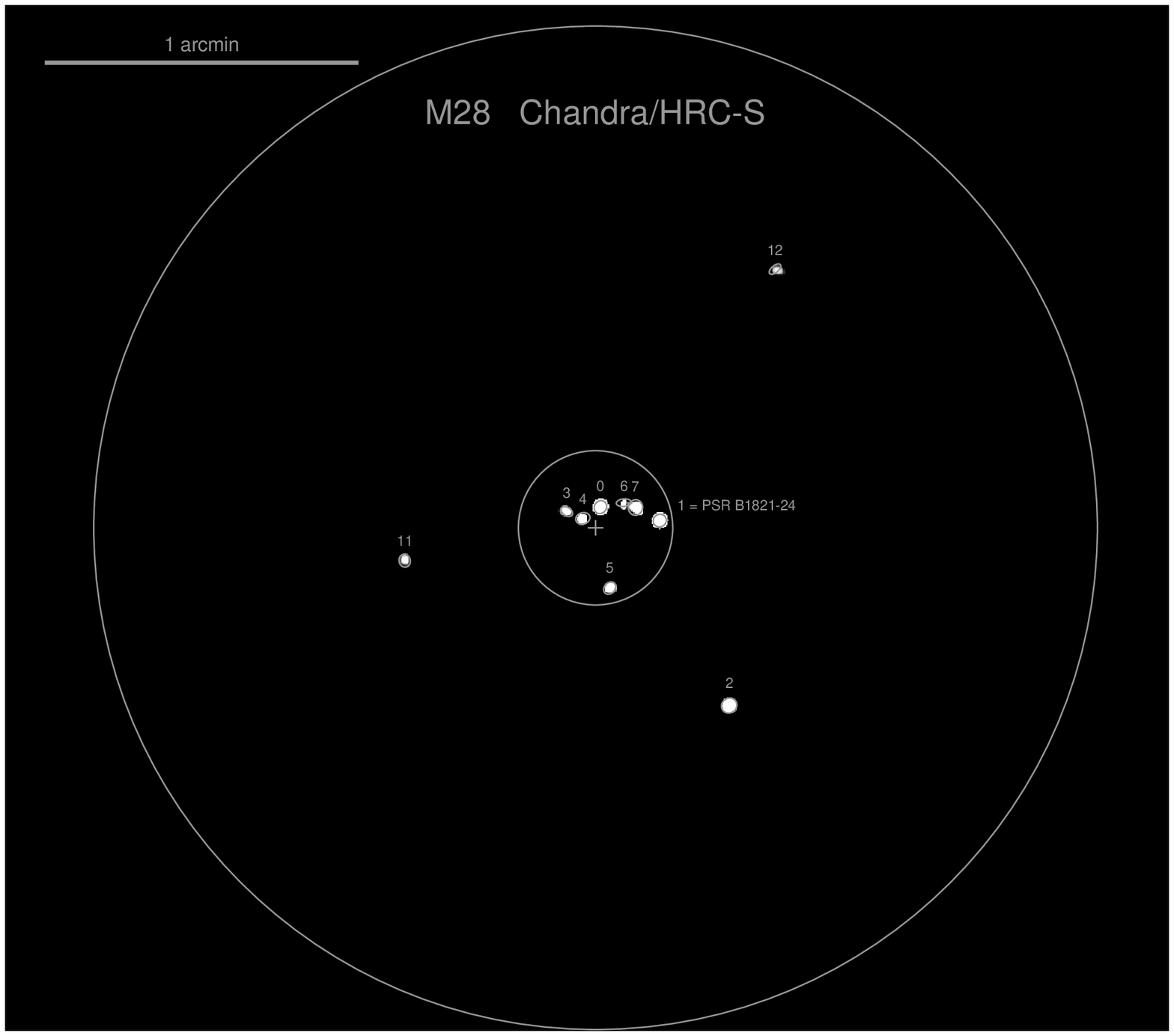 hoffset=-80 voffset=-80}{14.7cm}{21.5cm}
\FigNum{\ref{fig:wave}}
\end{figure}

\clearpage
\pagestyle{empty}
\begin{figure}[htb]
\PSbox{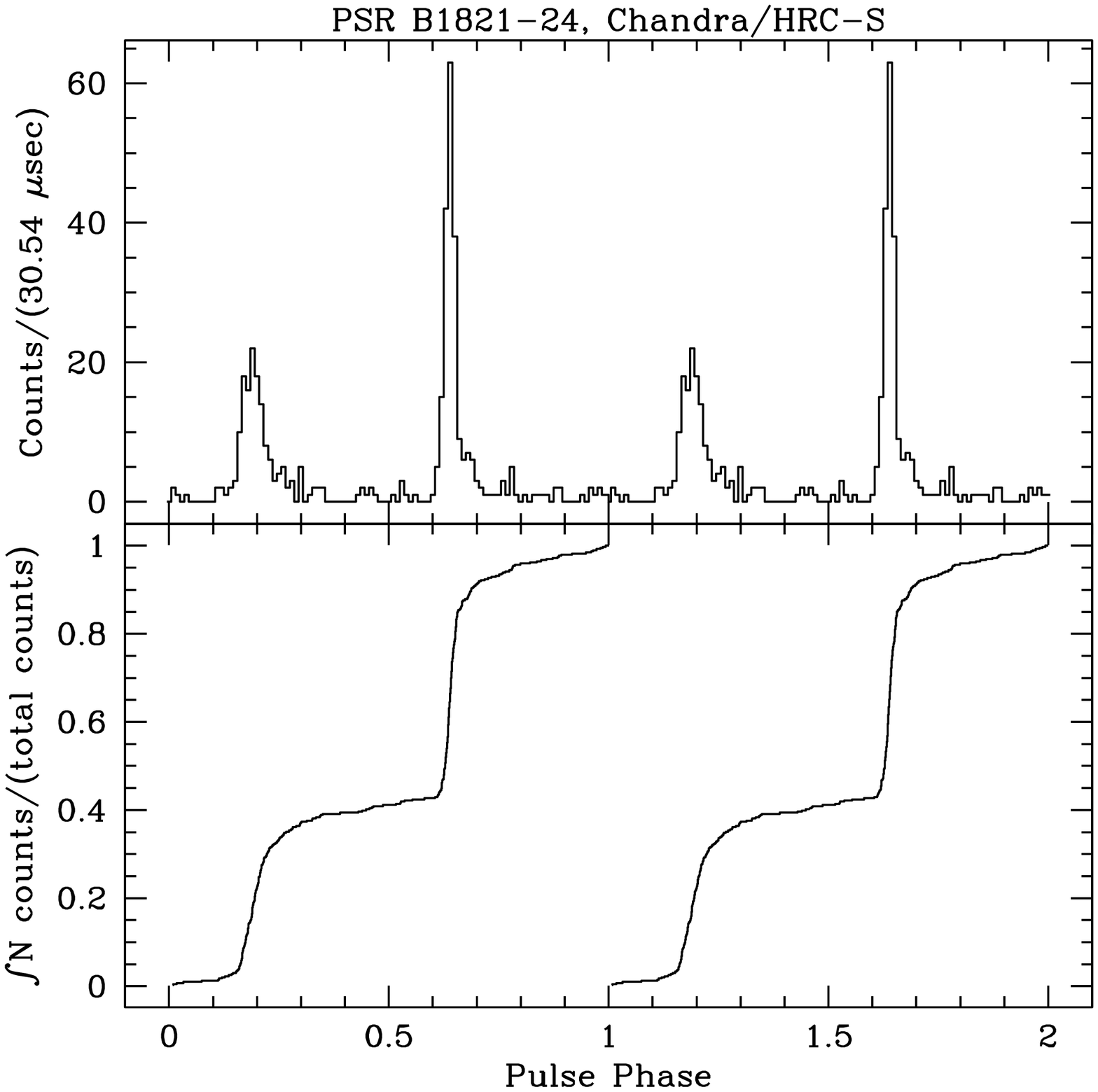 hoffset=-80 voffset=-80}{14.7cm}{21.5cm}
\FigNum{\ref{fig:phaselc}}
\end{figure}

\clearpage
\pagestyle{empty}
\begin{figure}[htb]
\PSbox{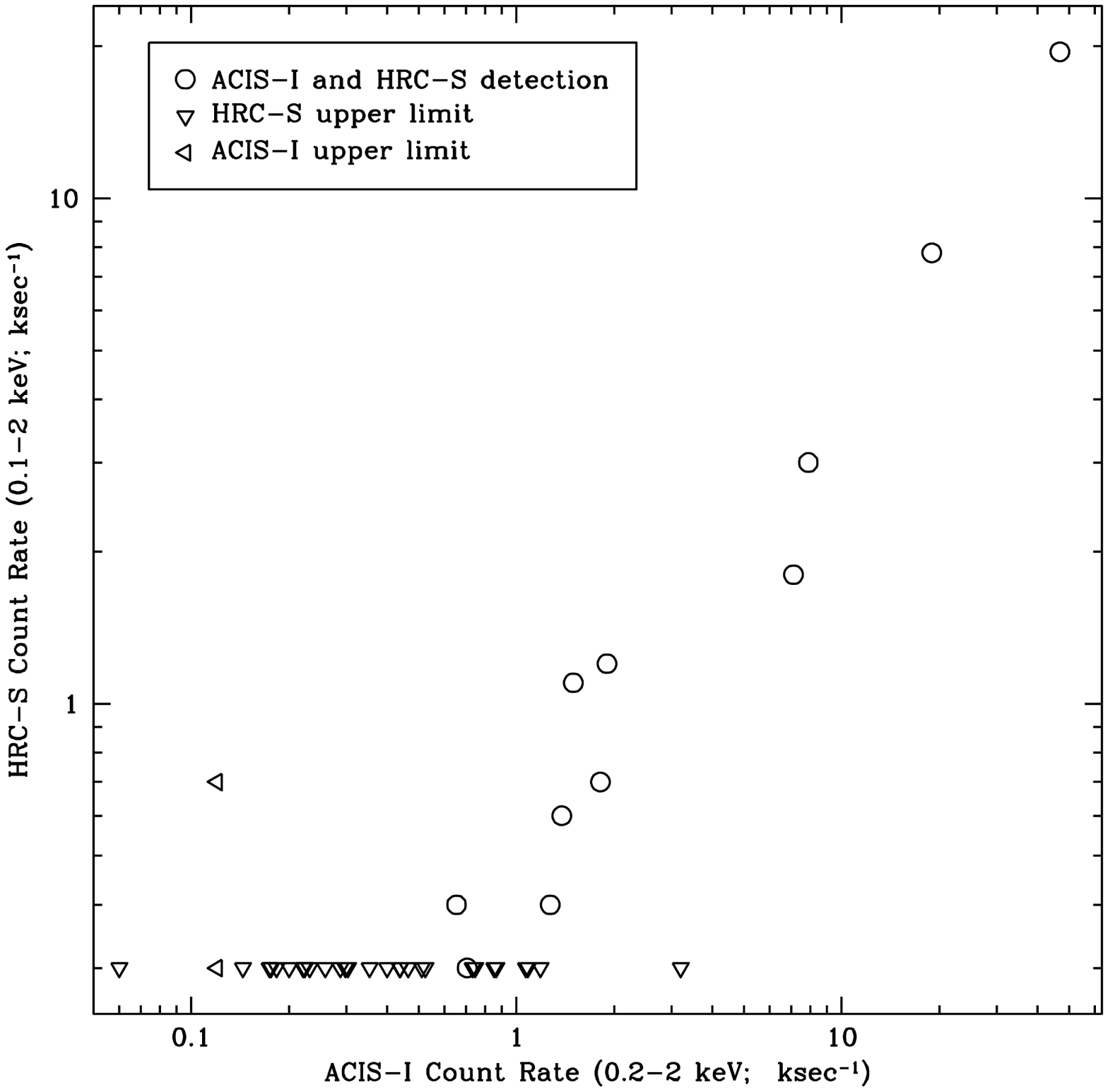 hoffset=-80 voffset=-80}{14.7cm}{21.5cm}
\FigNum{\ref{fig:cr}a}
\end{figure}

\clearpage
\pagestyle{empty}
\begin{figure}[htb]
\PSbox{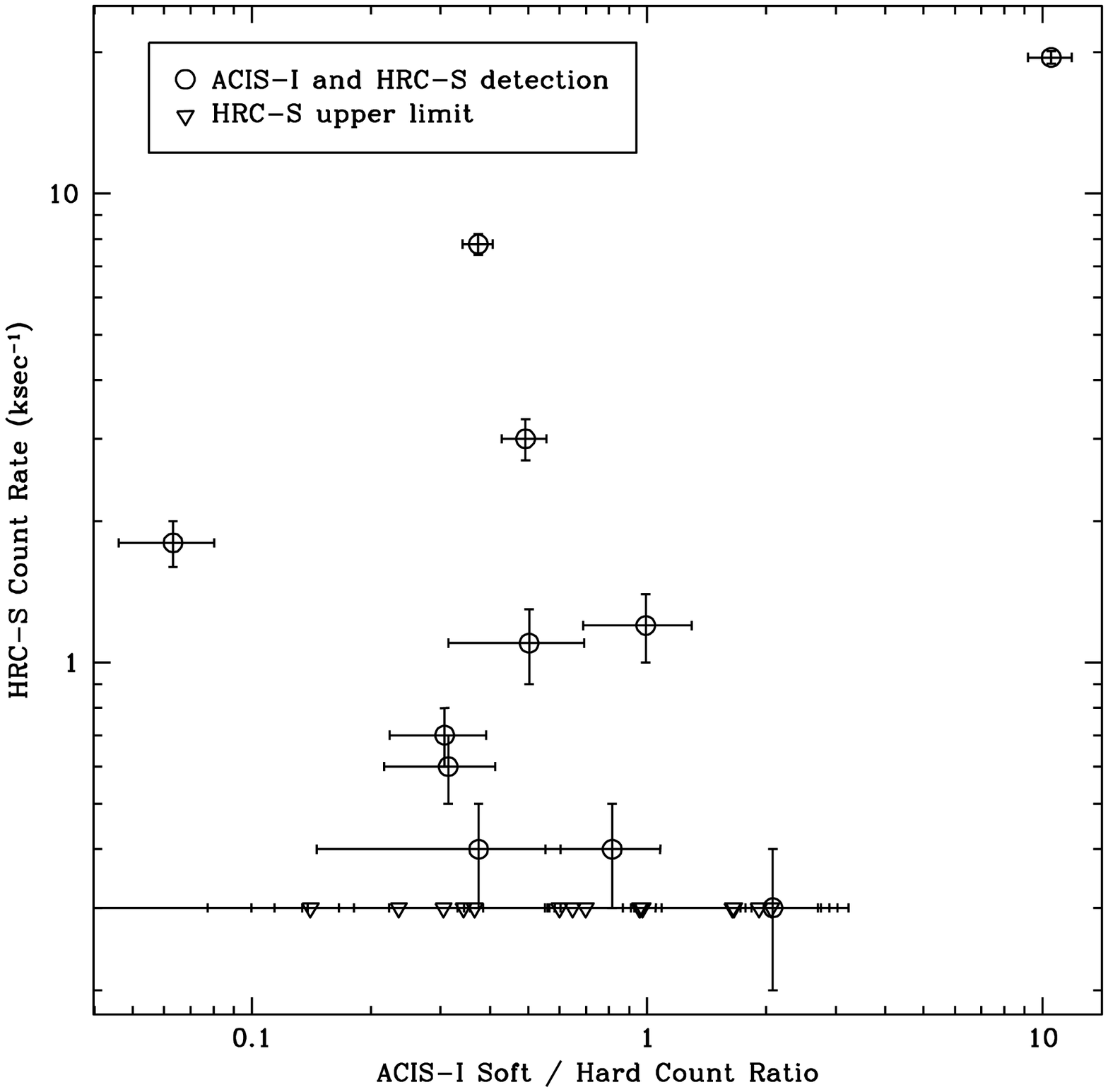 hoffset=-80 voffset=-80}{14.7cm}{21.5cm}
\FigNum{\ref{fig:cr}b}
\end{figure}

\clearpage
\pagestyle{empty}
\begin{figure}[htb]
\PSbox{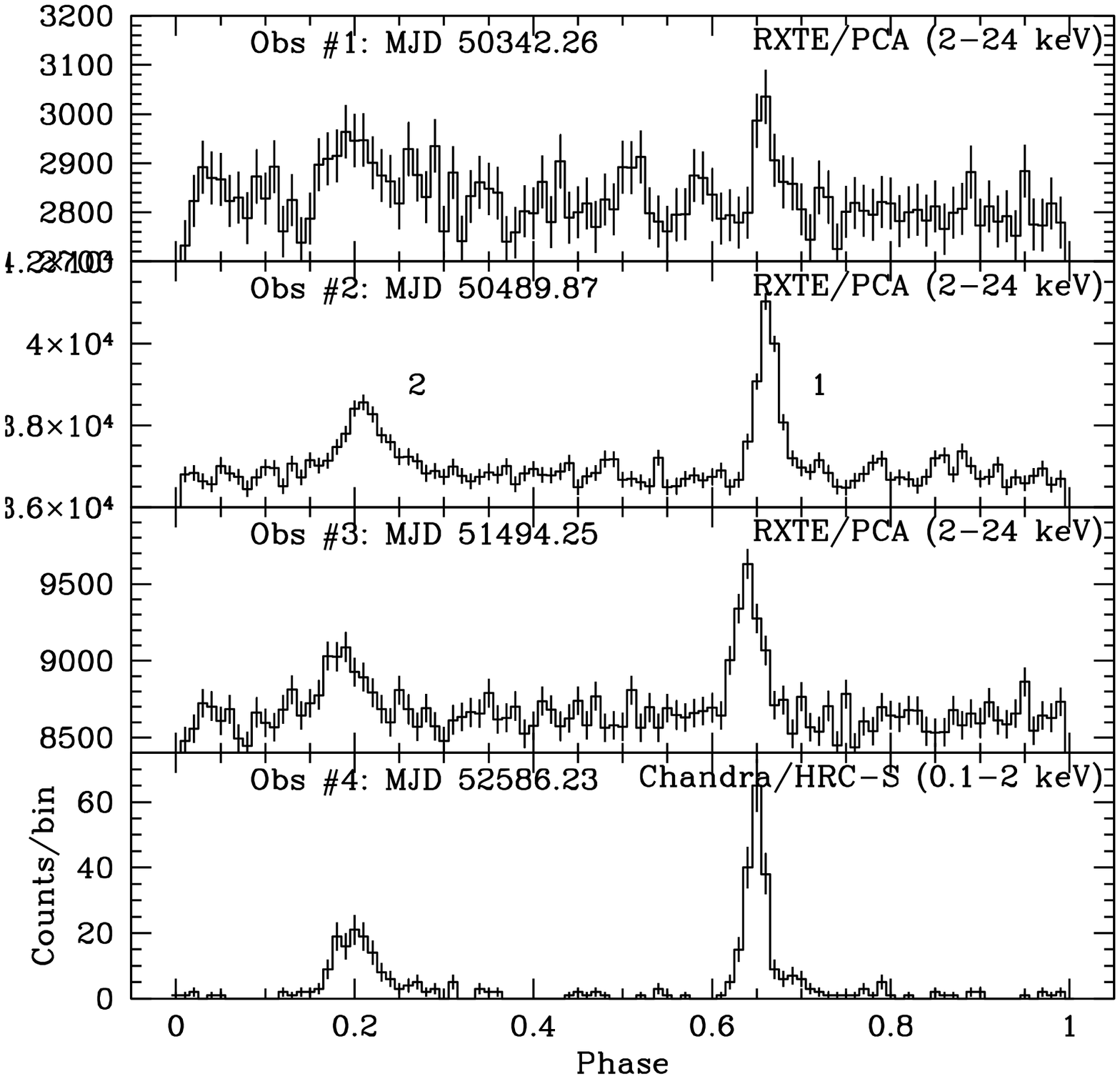 hoffset=-80 voffset=-80}{14.7cm}{21.5cm}
\FigNum{\ref{fig:allphases}}
\end{figure}

\clearpage
\pagestyle{empty}
\begin{figure}[htb]
\PSbox{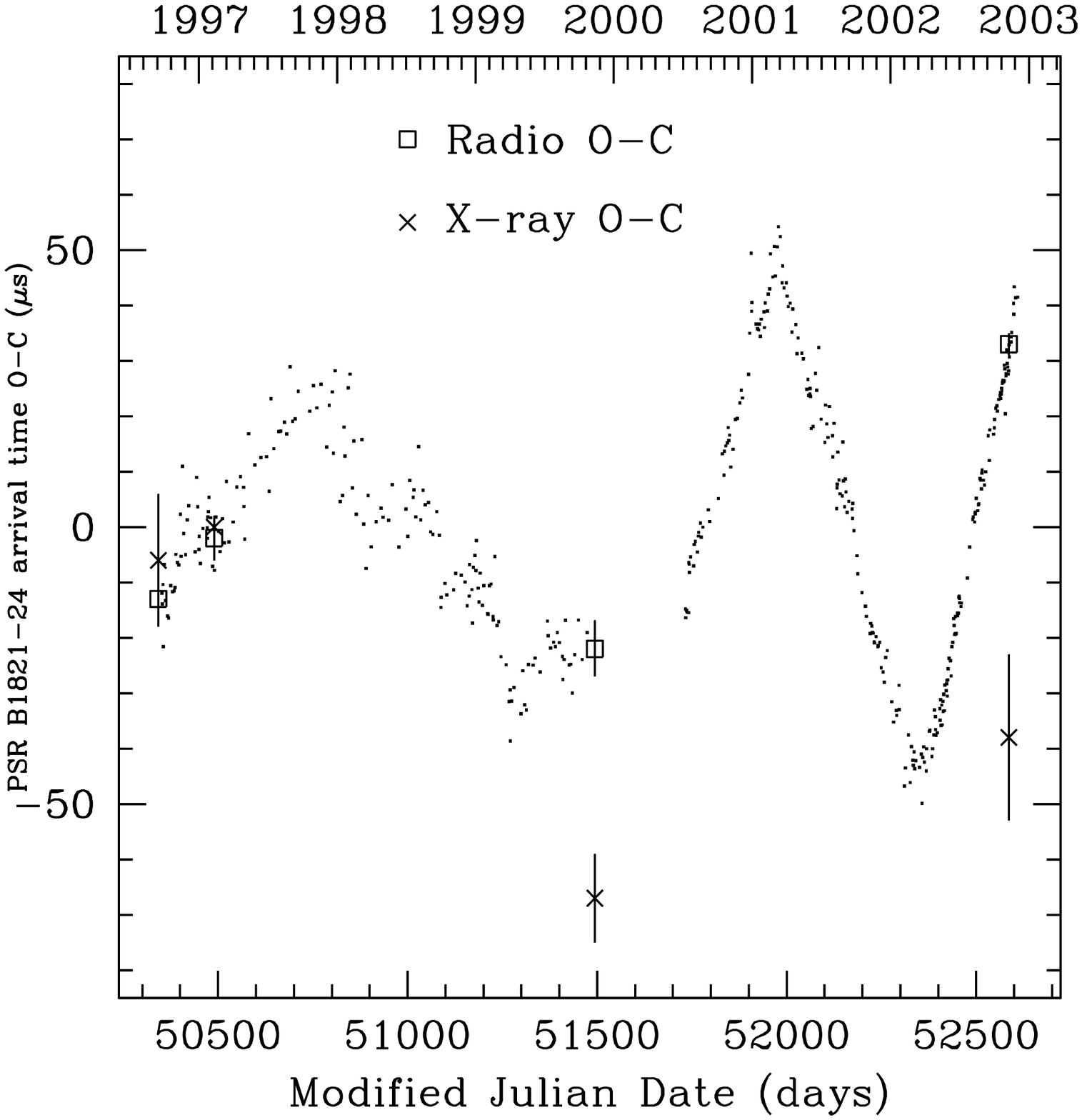 hoffset=-80 voffset=-80}{14.7cm}{21.5cm}
\FigNum{\ref{fig:timingnoise}}
\end{figure}

\clearpage
\pagestyle{empty}
\begin{figure}[htb]
\PSbox{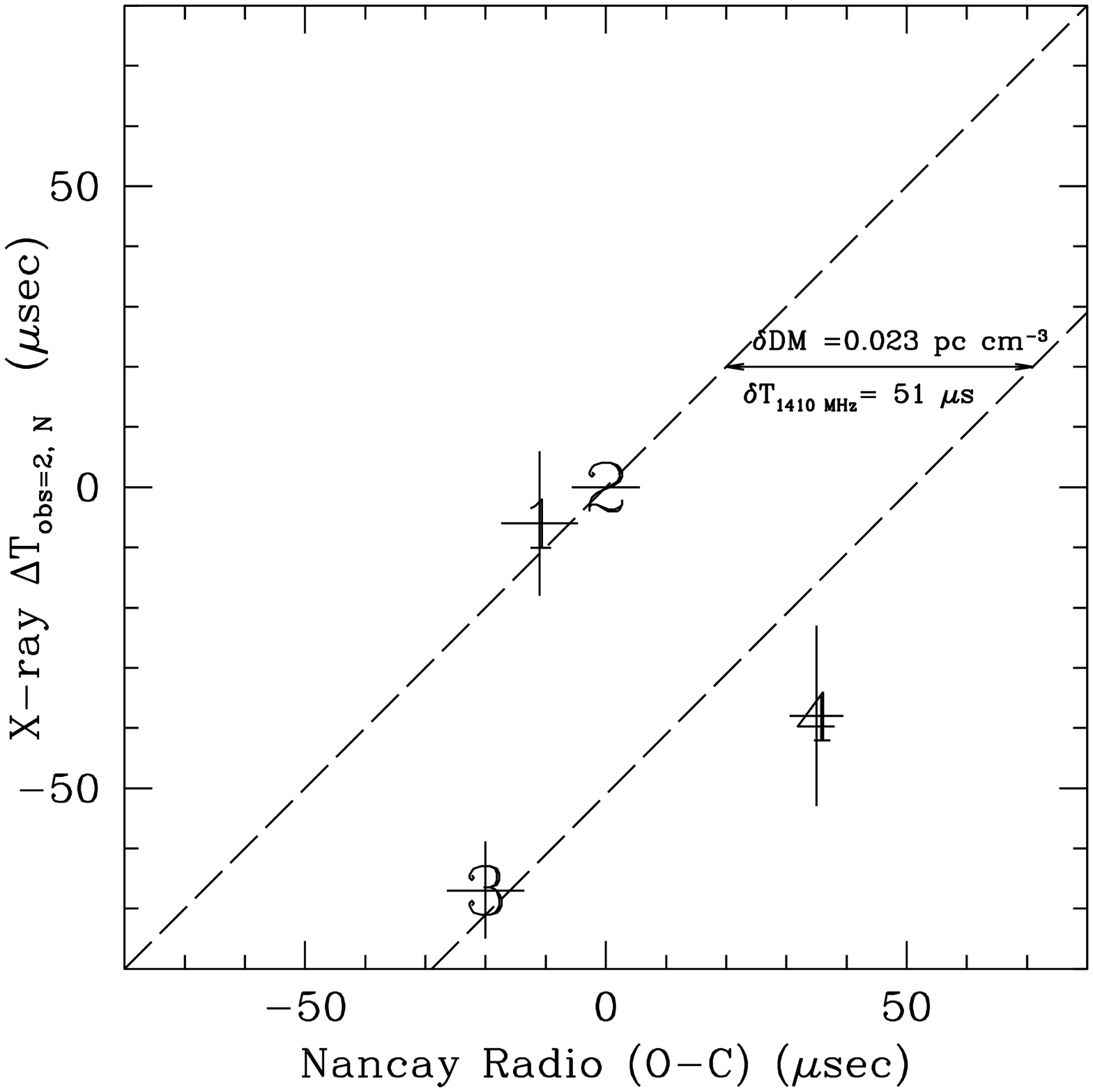 hoffset=-80 voffset=-80}{14.7cm}{21.5cm}
\FigNum{\ref{fig:money}}
\end{figure}

\begin{deluxetable}{lr}
\tablecaption{\label{tab:ephem}  \nancay\ Radio Ephemeris of \psr}
\tablewidth{5in}
\tablehead{
\colhead{Parameter} & 
\colhead{Value} \\
}
\startdata 
MJD Range				&  50351-52610		\\
$t_0$ (MJD)				&  51468.0		\\
R.A. (J2000)				&  18h24m32.s008345	\\
DEC.  (J2000)				&  $-$24d52m10.s758586	\\
DM (\dmunit)				&  119.873		\\
$\phi$($t_0$)				& $-$0.008(4) 		\\
$\nu$ (\perval{s}{-1})			& 327.40564101150(1)	\\
$\dot{\nu}$(\ee{-12} \perval{s}{-2})	&  $-$0.1735080(1)	\\
$\ddot{\nu}$ (\ee{-24}\perval{s}{-3})	&  0.66(1)		\\
RMS$^a$ (milliperiods)			& 8.1			\\
Time System				& DE405 		\\
\enddata
\tablecomments{The ephemeris is not corrected for the TOA variations
due to a time-variable DM, the effect of which is significant (that is,
comparable or greater than  the RMS variation observed).$\phi=0$ corresponds to the average arival phase of RP2 over the ephemeris MJD range.  
$^a$ root-mean-square deviations of Time of Arrival vs. ephemeris
prediction, in milli-periods.
}
\end{deluxetable}

\begin{deluxetable}{llccc}
\tablecaption{\label{tab:obs} Table of Observations}
\tablehead{
\colhead{Obs.}	& 
\colhead{Instrument } & 
\colhead{Obs. Start  }& 
\colhead{Obs. Start }& 
\colhead{Live Time )}\\
\colhead{ \#}	& 
\colhead{ (ObsID)} & 
\colhead{ (TT) }& 
\colhead{O (MJD) }& 
\colhead{ (ksec)}\\
}
\startdata
1& RXTE/PCA (P10421)	&1996 Sep 16 06:16	&50342.26&	6.6	\\
2& RXTE/PCA (P20159)	&1997 Feb 10 20:57	&50489.87&	98.6	\\
3& RXTE/PCA (P40090)	&1999 Nov 12 11:44	&51494.25& 	39.7  \\
4& \chandra/HRC-S (2797)  &2002 Nov 8  05:41   	&52586.23&	49.1 	\\
\enddata
\end{deluxetable}

\begin{deluxetable}{rrrrrrl}
\scriptsize
\tablecaption{\label{tab:catalog} Catalog of \chandra\ HRC-S X-ray Sources}
\tablehead{
\colhead{Label} & 
\colhead{RA (J2000.0)}& 
\colhead{Dec (J2000.0)}& 
\colhead{\ppm (\arcsec)} & 
\colhead{Dist. to core (\arcmin)} & 
\colhead{HRC-S \cperksec\ (\ppm) } & 
\colhead{\# from B02}
}
\startdata
          9  & 18 24 22.590  & -24 52  6.35  &  0.08 &  0.097 &    0.3 ( 0.1)&3   \\ 
          8  & 18 24 22.682  & -24 51  3.03  &  0.06 &  1.157 &    1.8 ( 0.2)&4   \\ 
         12  & 18 24 30.428  & -24 51 23.82  &  0.08 &  0.803 &    0.3 ( 0.1)&n/a \\ 
          2  & 18 24 31.056  & -24 52 45.28  &  0.04 &  0.555 &    3.0 ( 0.3)&17  \\ 
(\psr)    1  & 18 24 32.008  & -24 52 10.76  &  0.02 &  0.021 &    7.8 ( 0.4)&19  \\ 
          7  & 18 24 32.341  & -24 52  8.31  &  0.06 &  0.061 &    1.1 ( 0.2)&22  \\ 
          6  & 18 24 32.502  & -24 52  7.41  &  0.05 &  0.076 &    0.6 ( 0.1)&23  \\ 
          5  & 18 24 32.695  & -24 52 23.35  &  0.06 &  0.189 &    0.7 ( 0.1)&25  \\ 
          0  & 18 24 32.822  & -24 52  8.24  &  0.01 &  0.063 &   19.5 ( 0.6)&26  \\ 
          4  & 18 24 33.063  & -24 52 10.34  &  0.04 &  0.028 &    1.2 ( 0.2)&29  \\ 
          3  & 18 24 33.283  & -24 52  9.07  &  0.05 &  0.049 &    0.7 ( 0.1)&n/a \\ 
         11  & 18 24 35.506  & -24 52 18.25  &  0.08 &  0.105 &    0.4 ( 0.1)&37  \\ 
         10  & 18 24 36.591  & -24 50 15.71  &  0.09 &  1.939 &    0.4 ( 0.1)&38  \\ 
\enddata
\tablecomments{M28 Cluster center from \acite{harris96}. Astometry was
          performed relative to the \nancay\ ephemeris position (Table~\ref{tab:ephem})}
\end{deluxetable}

\begin{deluxetable}{rll}
\tablecaption{\label{tab:uncertainties} Sources and Magnitude of Absolute X-ray Timing Uncertainty}
\tablehead{
\colhead{Source} &
\colhead{Magnitude} \\
}
\startdata
\chandra\ HRC event digitization& 	\ppm 5 \musec  		\\
\chandra\  281\musec\ CCDM delay time	& 	\ppm 10 \musec  	\\
\chandra\ Clock Stability	& 	\ppm 5 \musec 		\\
RXTE/PCA event-UT tie	(after MJD 50567)	& 	\ppm 5 \musec  		\\	
RXTE/PCA event-UT tie	(before MJD 50567)	& 	\ppm 8 \musec  		\\	
\enddata
\end{deluxetable}

\begin{deluxetable}{ll}
\tablecaption{\label{tab:pulselc} X-ray Pulse Gaussian Model}
\tablewidth{5in}
\tablehead{
\colhead{Parameter} & 
\colhead{Value} \\
}
\startdata 
$\phi_1$ (phase)	& 0.6621 \\
$\sigma_1$ (phase)	& 0.0115\\
$A_1$			& 1.00\ppm0.08	\\
$\phi_2$ (phase)	& 0.2107\\
$\sigma_2$ (phase)	& 0.024\\
$A_2$			& 0.67\ppm0.07\\
\enddata
\tablecomments{Uncertainty in the relative normalization was found
from best-fit Gaussians to the \chandra\ data.}
\end{deluxetable}

\begin{deluxetable}{lllll}
\scriptsize
\tablecaption{\label{tab:values} Observational Timing Parameters}
\tablehead{
\colhead{Obs. } &
\colhead{Ref. Time} & 
\colhead{$\phi$(Ref. Time)} & 
\colhead{$\nu$(Ref.  Time)} & 
\colhead{$\dot\nu$(Ref. Time)} \\
\colhead{\#} &
\colhead{(MJD)} & 
\colhead{(frac.  phase)} & 
\colhead{(Hz)} & 
\colhead{(\ee{-12} \perval{s}{-2})} \\
}
\startdata
1 &50342.26240326679	&0.0480 &327.40565789065	&$-$0.173572	\\
2 &50489.86919495780	&0.3041 &327.40565567710	&$-$0.173564	\\
3 &51494.25419383658	&0.9761 &327.40564061792	&$-$0.173507	\\
4 &52586.23379831800	&0.0617 &327.40562425102	&$-$0.173444	\\
\enddata
\tablecomments{Values of $\phi$(Ref. Time), $\nu$(Ref. Time) and $\dot\nu$(Ref. Time) were
found using the DE405 time system ephemeris valid over MJD 50351-52610
(Table~\ref{tab:ephem}).  }
\end{deluxetable}

\begin{deluxetable}{llcllcllc}
\scriptsize
\tablecaption{\label{tab:phases} Phases of X-ray Components XP1 and
XP2, folded on \nancay\ Ephemeris}
\tablehead{
\colhead{Obs.} &
\colhead{$\phi_2$} & 
\colhead{$\frac{\phi_{2, \rm obs 2}-\phi_{2, \rm obs N}}{327.405 {\rm Hz}}$} & 
\colhead{$\sigma_2$} & 
\colhead{$\phi_1$} & 
\colhead{$\frac{\phi_{1, \rm obs 2}-\phi_{1, \rm obs N}}{327.405 {\rm Hz}}$} & 
\colhead{$\sigma_1$} & 
\colhead{X-ray $\Delta T_{obs=2,N}$} & 
\colhead{Radio (O$-$C)} 
\\
\colhead{N}	& 
\colhead{(phase)} &
\colhead{(\musec)} & 
\colhead{(phase)} & 
\colhead{(phase)} &
\colhead{(\musec)} & 
\colhead{(phase)} & 
\colhead{(\musec)} &
\colhead{(\musec)} \\
}
\startdata
1    		&0.199(6)	&37(20)		&[0.024]	&0.659(3)	&10(10)	&[0.0115]&  $-$6\ppm(8,8,5)  	& $-$13\ppm5 		\\
2 		&0.211(2)	&(0)		&0.025(3)	&0.6621(6)	&(0)	&0.0115(5)&  (0)	    	& $-2$\ppm4	\\
3 	&0.186(4)	&$-$76(14)	&0.023(5)	&0.640(1)	&$-$68(4)	&0.014(1) & $-$67\ppm(4,5,5)  & $-$22\ppm5		\\
4 	&0.202(3)	&$-$27(11)	&0.023(3)	&0.6490(9)	&$-$40(3)	&0.0115(9)&$-$38\ppm(4,12,5)& 33\ppm2		\\
\enddata
\tablecomments{The values in square brackets were held fixed. The values in parenthesis
are the uncertainty in the preceding digit(s). The uncertainties in
$\Delta T_{obs=2,N}$, (\musec) are: (statistical, systematic for
obs.\ N, systematic of obs.\ 2). }
\end{deluxetable}


\begin{thebibliography}{}

\bibitem[\protect\astroncite{{Backer} {\rm et~al.\/}}{1993}]{backer93}
{Backer}, D.~C., {Hama}, S., {van Hook}, S., \& {Foster}, R.~S., 1993,
\newblock {\em \apj} { 404}, 636

\bibitem[\protect\astroncite{{Backer} \& {Sallmen}}{1997}]{backer97}
{Backer}, D.~C. \& {Sallmen}, S.~T., 1997,
\newblock {\em \aj} { 114}, 1539

\bibitem[\protect\astroncite{{Becker} {\rm et~al.\/}}{2002}]{becker02}
{Becker}, W., {Swartz}, D., {Pavlov}, G., {Elsner}, R., {Grindlay}, J.,
  {Mignani}, R., {Tennant}, A., {Backer}, D., \& {Weisskopf}, M., 2002,
\newblock {\em \aap},
\newblock astro-ph/0211468

\bibitem[\protect\astroncite{{Becker} \& {Tr{\"u}mper}}{1999}]{becker99}
{Becker}, W. \& {Tr{\"u}mper}, J., 1999,
\newblock {\em \aap} { 341}, 803

\bibitem[\protect\astroncite{Bevington}{1969}]{bevington}
Bevington, P.~R., 1969,
\newblock {\em Data Reduction and Error Analysis for the Physical Sciences},
\newblock McGraw-Hill

\bibitem[\protect\astroncite{{Buccheri} {\rm et~al.\/}}{1983}]{buccheri+83}
{Buccheri}, R., {Bennett}, K., {Bignami}, G.~F., {Bloemen}, J.~B.~G.~M.,
  {Boriakoff}, V., {Caraveo}, P.~A., {Hermsen}, W., {Kanbach}, G.,
  {Manchester}, R.~N., {Masnou}, J.~L., {Mayer-Hasselwander}, H.~A., {Ozel},
  M.~E., {Paul}, J.~A., {Sacco}, B., {Scarsi}, L., \& {Strong}, A.~W., 1983,
\newblock {\em \aap} { 128}, 245

\bibitem[\protect\astroncite{{Chandra Science Center Flight Operations
  Team}}{2002}]{FOT02}
{Chandra Science Center Flight Operations Team}, 2002,
\newblock {\em Absolute Timing Accuracy},
\newblock Technical report, Center for Astrophysics

\bibitem[\protect\astroncite{{Cognard} {\rm et~al.\/}}{1996}]{cognard96}
{Cognard}, I., {Bourgois}, G., {Lestrade}, J.~., {Biraud}, F., {Aubry}, D.,
  {Darchy}, B., \& {Drouhin}, J.~., 1996,
\newblock {\em \aap} { 311}, 179

\bibitem[\protect\astroncite{{Cognard} {\rm et~al.\/}}{1995}]{cognard95}
{Cognard}, I., {Bourgois}, G., {Lestrade}, J.-F., {Biraud}, F., {Aubry}, D.,
  {Darchy}, B., \& {Drouhin}, J.-P., 1995,
\newblock {\em \aap} { 296}, 169

\bibitem[\protect\astroncite{{Cognard} \& {Lestrade}}{1997}]{cognard97}
{Cognard}, I. \& {Lestrade}, J.-F., 1997,
\newblock {\em \aap} { 323}, 211

\bibitem[\protect\astroncite{{Deshpande} {\rm et~al.\/}}{1992}]{deshpande92}
{Deshpande}, A.~A., {McCulloch}, P.~M., {Radhakrishnan}, V., \&
  {Anantharamaiah}, K.~R., 1992,
\newblock {\em \mnras} { 258}, 19P

\bibitem[\protect\astroncite{{Foster} {\rm et~al.\/}}{1988}]{foster88}
{Foster}, R.~S., {Backer}, D.~C., {Taylor}, J.~H., \& {Goss}, W.~M., 1988,
\newblock {\em \apjl} { 326}, L13

\bibitem[\protect\astroncite{{Foster} {\rm et~al.\/}}{1990}]{foster90}
{Foster}, R.~S., {Backer}, D.~C., \& {Wolszczan}, A., 1990,
\newblock {\em \apj} { 356}, 243

\bibitem[\protect\astroncite{{Foster} {\rm et~al.\/}}{1991}]{foster91}
{Foster}, R.~S., {Fairhead}, L., \& {Backer}, D.~C., 1991,
\newblock {\em \apj} { 378}, 687

\bibitem[\protect\astroncite{{Frail} {\rm et~al.\/}}{1991}]{frail91}
{Frail}, D.~A., {Cordes}, J.~M., {Hankins}, T.~H., \& {Weisberg}, J.~M., 1991,
\newblock {\em \apj} { 382}, 168

\bibitem[\protect\astroncite{{Frigo} \& {Johnson}}{1998}]{fftw}
{Frigo}, M. \& {Johnson}, S.~G., 1998,
\newblock in {\em ICASSP, Conference Proceedings, Vol. 3}, pp 1381--1384

\bibitem[\protect\astroncite{{Harris}}{1996}]{harris96}
{Harris}, W.~E., 1996,
\newblock {\em \aj} { 112}, 1487

\bibitem[\protect\astroncite{{HRC Team}}{1996}]{HRC96}
{HRC Team}, 1996,
\newblock {\em The High Resolution Camera (HRC) Description},
\newblock http://hea-www.harvard.edu/HRC/overview/overview.html

\bibitem[\protect\astroncite{{Leahy} {\rm et~al.\/}}{1983}]{leahy83}
{Leahy}, D.~A., {Darbro}, W., {Elsner}, R.~F., {Weisskopf}, M.~C., {Kahn}, S.,
  {Sutherland}, P.~G., \& {Grindlay}, J.~E., 1983,
\newblock {\em \apj} { 266}, 160

\bibitem[\protect\astroncite{{Lyne} {\rm et~al.\/}}{1987}]{lyne87}
{Lyne}, A.~G., {Brinklow}, A., {Middleditch}, J., {Kulkarni}, S.~R., \&
  {Backer}, D.~C., 1987,
\newblock {\em \nat} { 328}, 399

\bibitem[\protect\astroncite{{Murray} {\rm et~al.\/}}{1998}]{hrc1}
{Murray}, S.~S., {Chappell}, J.~H., {Kenter}, A.~T., {Kraft}, R.~P., {Meehan},
  G.~R., \& {Zombeck}, M.~V., 1998,
\newblock in {\em Proc. SPIE Vol. 3356, p. 974-984, Space Telescopes and
  Instruments V, Pierre Y. Bely; James B. Breckinridge; Eds.}, Vol. 3356, p.
  974

\bibitem[\protect\astroncite{{Murray} {\rm et~al.\/}}{2002}]{murray02}
{Murray}, S.~S., {Ransom}, S.~M., {Juda}, M., {Hwang}, U., \& {Holt}, S.~S.,
  2002,
\newblock {\em \apj} { 566}, 1039

\bibitem[\protect\astroncite{{Pavlov} {\rm et~al.\/}}{1999}]{pzt99}
{Pavlov}, G.~G., {Zavlin}, V.~E., \& {Tr{\" u}mper}, J., 1999,
\newblock {\em \apjl} { 511}, L45

\bibitem[\protect\astroncite{Press {\rm et~al.\/}}{1995}]{press}
Press, W., Flannery, B., Teukolsky, S., \& Vetterling, W., 1995,
\newblock {\em Numerical Recipies in C},
\newblock Cambridge University Press

\bibitem[\protect\astroncite{{Ransom} {\rm et~al.\/}}{2002}]{rem02}
{Ransom}, S.~M., {Eikenberry}, S.~S., \& {Middleditch}, J., 2002,
\newblock {\em \aj} { 124}, 1788

\bibitem[\protect\astroncite{{Rots} {\rm et~al.\/}}{1998}]{rots98}
{Rots}, A.~H., {Jahoda}, K., {Macomb}, D.~J., {Kawai}, N., {Saito}, Y.,
  {Kaspi}, V.~M., {Lyne}, A.~G., {Manchester}, R.~N., {Backer}, D.~C., {Somer},
  A.~L., {Marsden}, D., \& {Rothschild}, R.~E., 1998,
\newblock {\em \apj} { 501}, 749

\bibitem[\protect\astroncite{{Saito} {\rm et~al.\/}}{1997}]{saito97}
{Saito}, Y., {Kawai}, N., {Kamae}, T., {Shibata}, S., {Dotani}, T., \&
  {Kulkarni}, S.~R., 1997,
\newblock {\em \apjl} { 477}, L37

\bibitem[\protect\astroncite{Watson \& Davis}{2002}]{Watson02}
Watson, M. \& Davis, W., 2002,
\newblock {\em Chandra Clock Correlation Summary},
\newblock Technical report, Center for Astrophysics

\bibitem[\protect\astroncite{{Wijnands} \& {Van der Klis}}{1998}]{rudi98}
{Wijnands}, R. \& {Van der Klis}, M., 1998,
\newblock {\em \nat} { 394}, 344

\bibitem[\protect\astroncite{{Zombeck} {\rm et~al.\/}}{1995}]{hrc2}
{Zombeck}, M.~V., {Chappell}, J.~H., {Kenter}, A.~T., {Moore}, R.~W., {Murray},
  S.~S., {Fraser}, G.~W., \& {Serio}, S., 1995,
\newblock in {\em Proc. SPIE Vol. 2518, p. 96-106, EUV, X-Ray, and Gamma-Ray
  Instrumentation for Astronomy VI, Oswald H. Siegmund; John V. Vallerga;
  Eds.}, Vol. 2518, p.~96

\end{thebibliography}
\end{document}